\documentclass[journal]{IEEEtran}
\IEEEpeerreviewmaketitle

\usepackage{cite}
\usepackage{amsmath,amssymb,amsfonts}
\usepackage{algorithmic}
\usepackage{graphicx}
\usepackage{textcomp}
\usepackage{color,soul} 
\usepackage{subfigure}
\usepackage{threeparttable}
\usepackage{float}
\usepackage{epsfig}
\usepackage{epstopdf}
\usepackage{amsmath}
\usepackage{multirow}
\usepackage{makecell}

\begin{document}

\title{\LARGE{Impact of Reconfigurable Intelligent Surface Geometry on Communication Performance}}

\author{Zhuangzhuang Cui,~\IEEEmembership{Member,~IEEE},~and~Sofie Pollin,~\IEEEmembership{Senior Member,~IEEE} 
\thanks{
This work is supported by European Union's Horizon 2020 research and innovation program under grant no. 101013425 (REINDEER), and KU Leuven Postdoctoral Mandate (PDM) under project no. 3E220691.}
\thanks{Z. Cui and S. Pollin are with the WaveCoRE of Department of Electrical Engineering (ESAT), KU Leuven, 3000 Leuven, Belgium (E-mail: \{zhuangzhuang.cui, sofie.pollin\}@kuleuven.be).}}

\maketitle
\begin{abstract}
When beamforming is applied at the transmitter, only part of reconfigurable intelligent surfaces (RISs) will be active, and it becomes indispensable to study the impact of RIS geometry. This letter aims at evaluating RIS geometry, ranging from linear (1D), and planar (2D), to cylindrical (3D) structures. We first derive the effective illuminated surface of different RIS topologies and determine the resulting signal-to-noise ratio (SNR) and outage probability. Then, we investigate the optimal RIS location to trade off active area versus received power considering near-field propagation. Numerical results quantify the benefit of RIS geometric compactness and provide their applicable ranges.
\end{abstract}


\IEEEpeerreviewmaketitle
\section{Introduction}
Reconfigurable intelligent surface (RIS) technology is attracting much interest for the sixth-generation (6G) communication systems. It is believed to be an essential component when blockages occur between transmitter (Tx) and receiver (Rx) \cite{Strinati21}. 
Passive RIS furthermore has the potential to combine high signal-to-noise ratio (SNR) with high energy efficiency. 
Existing studies have unfolded the performance analysis of RIS-assisted wireless communication. In general, these works can be categorized into three kinds: \textit{i}) from the perspective of propagation channels, the links between the Tx and RIS (TS), Rx and RIS (RS) are assumed to follow Rayleigh \cite{cui21}, Rician \cite{salhab21}, or Nakagami-$m$ \cite{hhu21} distributions, mostly assuming the far-field propagation; \textit{ii}) from the perspective of frequency bands, the operating band spans from sub-6~GHz \cite{dai20} to millimeter-wave (mmWave) bands \cite{basar21}; \textit{iii}) from the perspective of application scenarios, the deployments of RIS spread over the cellular system \cite{JQP22}, and vehicular system \cite{ammc22}, mainly for outdoor environments. 
There is a lack of studies focusing on indoor deployments with near-field effects.

The dominant RIS geometry assumption in existing studies is the two-dimensional (2D) plane \cite{huang19}, which is the natural interpretation of a surface. 
Another hypothesis is that \textit{all} elements in a RIS are 
actively reflecting the incoming wave, and contribute to the SNR gain \cite{basar19}. Such assumptions might hold in far-field scenarios, it is not effective when the RIS is only partially illuminated due to the large aperture, the short distance between the transceiver and the RIS in an indoor environment, or a high-directional beam pattern used in Tx. When in the near field the beam is focused locally, the impact of the RIS compactness might be important. Thus, the performance benefit of the 2D surface, by comparison with other geometries, such as linear (1D), and cylindrical (3D) ones, needs more analysis. Therefore, this paper aims to evaluate the superiority of different RIS geometries.

Nonetheless, a few studies have investigated the partially illuminated RIS. In \cite{Ntontin2021}, the authors approximated the illuminated RIS area 
from the conical main lobe as a circular area, and then determined the effective number of RIS elements. Their extended work in \cite{Ntontin2022} considered an ellipse as the illuminated area by incorporating an incident angle. Authors in \cite{GSA2022} investigated the RIS efficiency indicated by the received power for different user directions, 
while modeling the partial RIS illumination 
as an elliptical area 
that degrades 
to a circle in the normal direction. However, these works merely focused on 2D RIS and studied the deployment of RIS with limited flexibility. In our work, the illuminated area is a function of the half-power beam-width (HPBW), the distance, as well as both azimuth and elevation of departure (AoD and EoD). To our best knowledge, this is the first study that investigates the impact of different RIS geometries on communication performance considering more practical 3D deployment of RIS. The main contributions are as follows.
\begin{itemize}
    \item We theoretically formulated the illuminated area and the effective number of different RIS geometries, as functions of the distance, the HPBW, and the angle of departure.
    \item We considered the near-field effects in the path loss and obtained near-field distances for different RIS geometries. We then derived the outage probability, by means of the moment matching method, which can be used for any effective number of RIS elements.  
    \item We compared the effective number of 
    elements, received power, SNR, and outage probability for different geometries, to provide practical deployment 
    insights. Moreover, the optimal location of RIS was also investigated to realize the optimal SNR and the minimum outage probability.
\end{itemize}

The remainder of the paper is structured as follows. Section II will introduce the system model, and illuminated area when using beamforming in Tx. Section III will analyze the effective number of elements for different RIS geometric structures, and then provide the expression of outage probability. Afterward, Section IV will provide analytical results and deployment insights into RISs. Finally, Section V will conclude the paper.

\section{System Model}
In this section, we will introduce the system model, and model SNR as a function of signal transmission, antenna pattern, and channel model. In addition, we will calculate the illuminated area, which plays an important role in determining the effective number of elements and near-field distance. 


\subsection{Signal Transmission}
For the system model in Fig.~1, we consider the unfavorable propagation condition, i.e., an absent Line-of-Sight (LoS) path between Tx and Rx, where the RIS can be beneficial to compensate for the severe propagation loss. At the Rx, the received signal can be expressed as
\begin{equation}
    y=\sqrt{P_r}\left[\sum_{i=1}^{N_{\rm eff}}h_i \varrho_i g_i\right]s+n_0,
    \label{signal_trans}
\end{equation}
where $P_r$ is the received power. $h_i=\alpha_i\exp(-j\varpi_i)$ and $g_i=\beta_i\exp(-j\vartheta_i)$ represent the small-scale fading channel for Tx-RIS and RIS-Rx links, where $\alpha_i$ and $\beta_i$ follow Rayleigh distribution with parameters $\sigma_1$ and $\sigma_2$, respectively. Moreover, $\varrho_i$ is the reflection coefficient in the $i$-th RIS element. $s$ and $n_0$ represent the transmit signal with $\mathbb{E}(s^2)=1$ and the additive white Gaussian noise (AWGN) with $n_0\sim \mathcal{N}(0,\sigma_n^2)$. Finally, $N_{\rm eff}$ is the effective number of RIS elements. 



\subsection{Antenna and Illuminated Area}
For a large RIS, existing studies show that only partial elements will be illuminated when the Tx uses 
beamforming. Herein, we denote the HPBW as $\phi$, and based on the cosine model \cite{yzh17}, the Tx antenna pattern for antenna element $n$ can be expressed as 
$G_t(\varphi_{t,n})=\cos^2\left(\frac{\pi(\varphi_{t,n}-\varphi_t)}{2\phi}\right)$, with $\varphi_t$ denoting the azimuth angle between Tx and the center of RIS. For near-field conditions, $\varphi_{t,n}$ is diverse for different $n$. 
The beamforming is assumed to be a simple phased array with steering vector pointing to the center of the RIS structure.
Then, the element radiation pattern of the RIS is expressed as $F(\varphi_{t,n})=\cos^3(\varphi_{t,n})$ for $\varphi_{t,n}\sim[-\frac{\pi}{2},\frac{\pi}{2}]$ and $\theta_{t,n}\sim[-\frac{\pi}{2},\frac{\pi}{2}]$ where $\varphi_{t,n}, \theta_{t,n}$ denote the AoD and EoD for each element, respectively. The same pattern applied for $\varphi_{r,n}, \theta_{r,n}$. 
Finally, the Omni-directional antenna pattern applies in the Rx. 

For the illuminated area, we approximate it as an ellipse, and its major axis is denoted by $a$, which can be expressed as
\begin{equation}
    a=\frac{1}{2}\left(\frac{r_1\sin(\phi_0)}{\sin(\varphi_t+\phi_0)}+\frac{r_1\sin(\phi_0)}{\sin(\varphi_t-\phi_0)}\right),
    \label{ellipse_a}
\end{equation}
where $a$ can be obtained by applying the Law of Sines, $\varphi_t$ represents the AoD, and $r_1$ is the 3D distance between the Tx and RIS. Similarly, the minor axis $b$ can be calculated by
\begin{equation}
    b=\frac{1}{2}\left(\frac{r_1\sin(\phi_0)}{\cos(\theta_t+\phi_0)}+\frac{r_1\sin(\phi_0)}{\cos(\theta_t-\phi_0)}\right),
    \label{ellipse_b}
\end{equation}
where $\theta_t=\arctan{\frac{h_s-h_t}{d_1}}$ is the EoD and $d_1$ is 2D distance between Tx and RIS ($d_2$ for RIS-Rx). The detailed explanation and calculation of $a$ and $b$ can be found in Appendix A.

Therefore, the illuminated area on the 2D plane is calculated by $S_{\rm m}=\pi a b$
where the area becomes a sphere ($a=b$) when $\theta_t=0^{\circ}, \varphi_t=90^{\circ}$. Moreover, for $\theta_t=90^{\circ}$ or $\varphi_t=0^{\circ}$, it corresponds to the beam parallel to the RIS, thus $S_{\rm m}=0$.
\subsection{Path Loss and SNR}
For path loss, we employ the measurement results from \cite{tang21}. The received power considering the near field is calculated by 
\begin{equation}
    P_r=P_t G_r\frac{\lambda^2 d_x d_y}{64 \pi^3}\left|\sum_{n=1}^{N_{\rm eff}}\frac{\sqrt{G_t (\varphi_{t,n})F(\varphi_{t,n})F(\varphi_{r,n})}}{r_{1,n} r_{2,n}}\right|^2,
    \label{pr_near}
\end{equation}
where $P_t$ is the transmit power. $d_x$$\times$$d_y$ represents the size of an element, often assumed to be sub-wavelength. $r_{1,n}$ and $r_{2,n}$ represent 3D distances of Tx-element and element-Rx, respectively. 
The distances 
$r_{1,n},r_{2,n}$ and angles $\varphi_{t,n},\varphi_{r,n}$ are different for each RIS element ($n=1...N_{\rm eff}$) in the near field. When operating in the far field, the plane wave assumption still results in different $r_{1,n}$ and $r_{2,n}$ in the 3D structure. 

Thus, the instantaneous SNR can be expressed by
\begin{equation}
    \rho_{\rm inst}=\frac{P_r|\sum_{i=1}^{N_{\rm eff}}\alpha_i\beta_i\exp(\psi_i-\varpi_i-\vartheta_i)|^2}{\sigma_n^2},
    \label{signal_trans}
\end{equation}
where the 
channel phase shift can be perfectly compensated by the RIS, i.e., $\psi_i=\varpi_i+\vartheta_i$. Hence, $\rho_{\max}=A^2\bar{\rho}$ with $A=\sum_{i=1}^{N_{\rm eff}}\alpha_i\beta_i$, and $\bar{\rho}=\frac{P_r}{\sigma_n^2}$ denoting the average SNR.

\begin{figure}[!t]
  \centering
   {\includegraphics[width=3in]{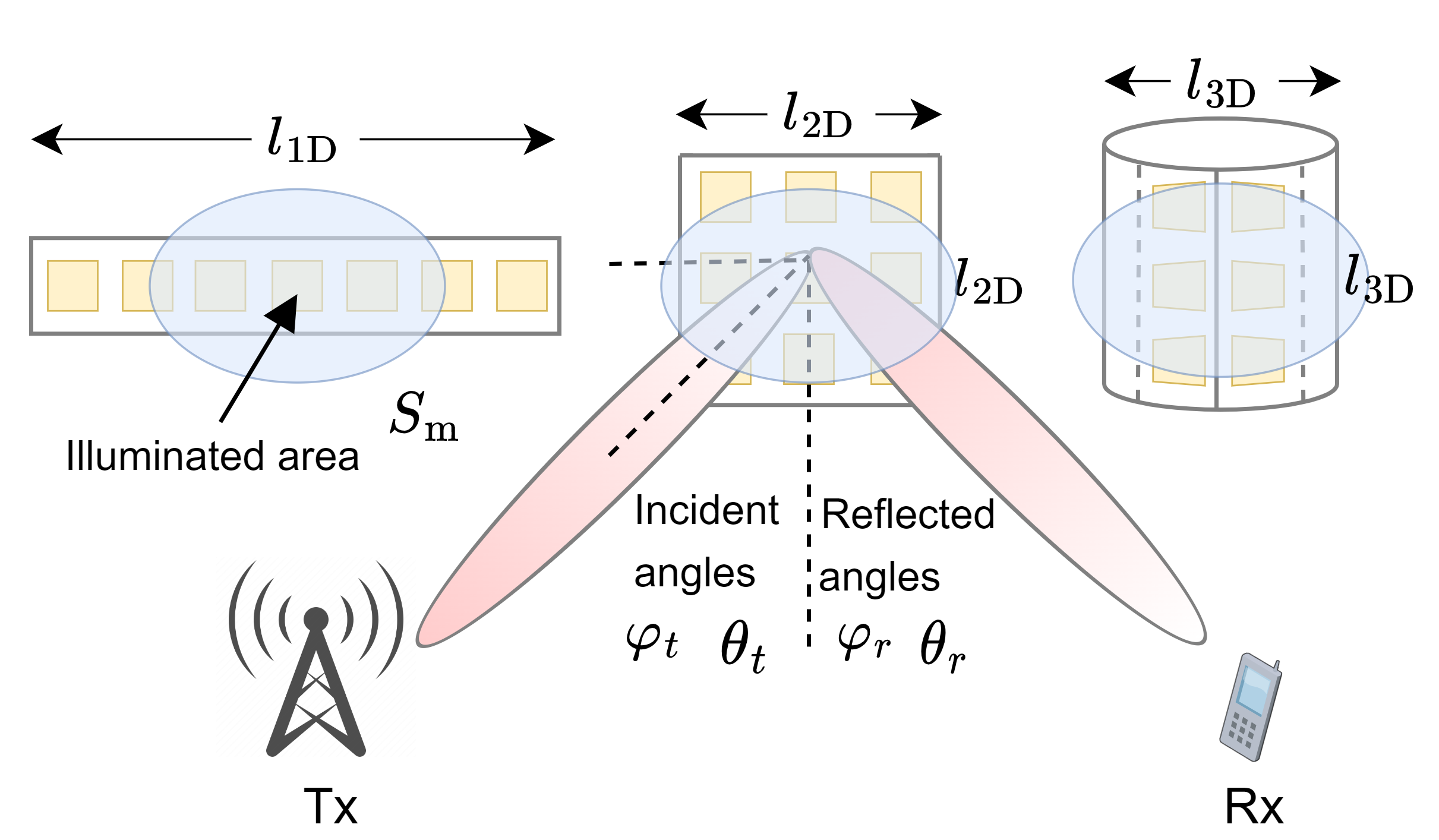}}
  \caption{An illustration of RIS geometric structures, where the illuminated area is $S_{\rm m}$, and azimuth and elevation angles are $\varphi_t$, $\varphi_r$ and $\theta_t$, $\theta_r$, respectively.}
  \label{models}
 \end{figure}



\section{Key Issues and Analysis}

In this section, we will first address key issues on the effective number $N_{\rm eff}$, the Fraunhofer distance $d_{G, G\sim{\rm\{1D, 2D, 3D}\}}$, and element-dependent distance $r_{1,n}$ and $r_{2,n}$.
Afterward, we will derive the outage probability as function of the SNR. 


\subsection{Effective Number of RIS Elements}

The effective number of elements is determined by the illuminated area, however, the calculation is different for diverse RIS geometries. For the linear RIS, the effective number is mainly determined by the major axis $a$ because of $b\ge d_y$, expressed as $N_{\rm eff}^{\rm 1D}=\left\lfloor\frac{a}{d_s}\right\rfloor$
where we use $\lfloor \cdot \rfloor$ to disregard the partially illuminated element. For a 2D squared RIS, its length can be expressed as $l_{\rm 2D}=(2\sqrt{N}-1)d_x$. To simplify the scenario analysis, we introduce a parameter $I_{\rm 2D}$ to indicate the relationship between the elliptical illumination and squared RIS areas. We compute $I_{\rm 2D}=(2a-l_{\rm 2D})(2b-l_{\rm 2D})$, where $I_{\rm 2D} > 0$ 
means that the RIS is either fully illuminated (the ellipse dimensions $a$ and $b$ are both larger than $l_{2D}$ - and the area is given by $S_s$) or either partially illuminated in both dimensions so the area is given by the ellipse area $S_{\rm m}$. Under this condition, the effective number can be calculated by
\begin{equation}
N_{\rm eff}^{\rm 2D}=
\begin{cases}
N, & S_{\rm m}\ge S_s, I_{\rm 2D} > 0,\\
\left\lfloor \frac{S_{\rm m}}{2d_s^2} \right\rfloor, &  S_{\rm m} < S_s, I_{\rm 2D} > 0,\\
\end{cases}
\end{equation}
where $S_s$ is the total area of the RIS. For the scenarios where $I_{\rm 2D} \le 0$, the RIS is partially illuminated in one dimension and fully in the other. 
As $a \ge b$, $I_{\rm 2D} = 0$ holds when $b=l_{\rm 2D}/2$, where $N_{\rm 2D} =\lfloor S_{\rm m}/2d_s^2 \rfloor$. For the other condition, the number is calculated as $N_{\rm eff}^{\rm 2D}=\left\lfloor\frac{S_{\rm m}-S_1}{2d_s^2} \right\rfloor$ for $a > \frac{l_{\rm 2D}}{2}, I_{\rm 2D} < 0$.
where the geometric illustration of $S_1$ is shown in Fig.~A1(b). 
Based on the elliptic Equation, $S_1$ can be calculated as 
\begin{equation}
\begin{aligned}
S_1&=\frac{4b}{a}\int_{\frac{l_{\rm 2D}}{2}}^{a}\sqrt{a^2-x^2}dx,\\
&=\pi ab-\frac{bl_{\rm 2D}}{2a}\sqrt{4a^2-l_{\rm 2D}^2}-2ab\arcsin{\frac{l_{\rm 2D}}{2a}}.
\end{aligned}
\end{equation}

For a 3D cylindrical RIS, the height is calculated as $l_{\rm 3D}=\sqrt{\frac{(d_xd_y+d_s^2)N}{\pi}}$. The effective number can be determined according to the relation between the illuminated and cylindrical surface. However, due to the nature of the cylinder, the mapping major axis of the ellipse on the cylinder is 
\begin{equation}
    a^{*}=\frac{\pi l_{\rm 3D}}{4}-\frac{l_{\rm 3D}}{2}\arccos\left(\frac{2a}{l_{\rm 3D}}\right), \; a \le \frac{l_{\rm 3D}}{2},
\end{equation}
where $a^{*}$ is the arc length in a circle when chord length is $a$.
 \begin{figure}[!t]
  \centering
   {\includegraphics[width=2.5in]{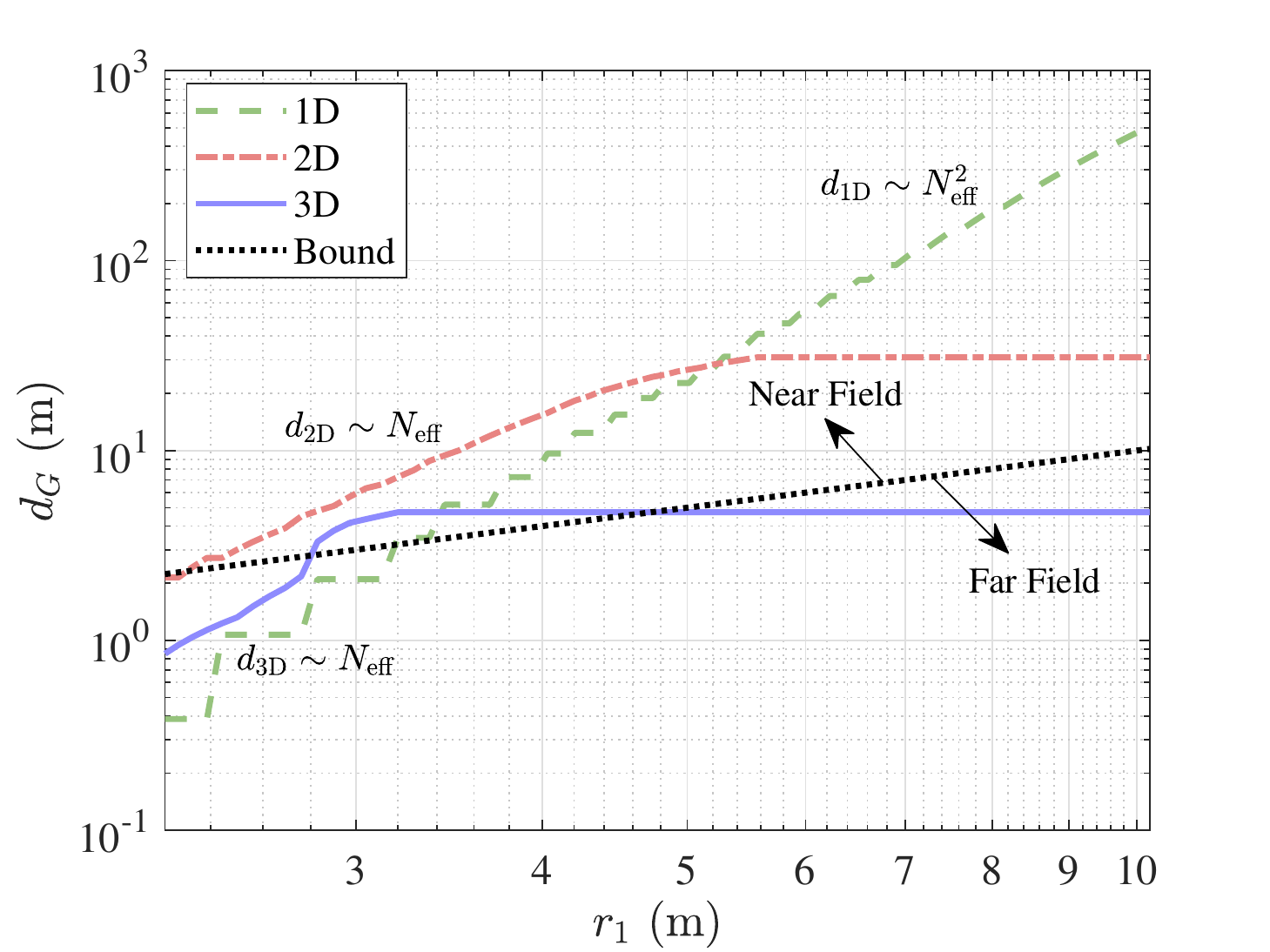}}
 \caption{$d_G$ vs. $r_1$ for RIS geometries ($\phi=5^\circ$), where $d_{\rm 1D}$ increases faster than $d_{\rm 2D, 3D}$, and $d_{G}$ becomes constant when $N_{\rm eff}^{G}$ reaches the limit.}
 \label{far-field-distance}
\end{figure} 

In summary, the effective number of the 3D RIS for different ranges of $a$ and $b$ can be expressed as
\begin{equation}
N_{\rm eff}^{\rm 3D}=
\begin{cases}
\left\lfloor \frac{S_2+S_3}{2d_s^2} \right\rfloor, & a > \frac{l_{\rm 3D}}{2}, b < \frac{l_{\rm 3D}}{2}\\
\frac{N}{2}, & a > \frac{l_{\rm 3D}}{2}, b > \frac{l_{\rm 3D}}{2}\\
\left\lfloor \frac{\pi a^{*} b}{2d_s^2} \right\rfloor, & a \le \frac{l_{\rm 3D}}{2}, b \le \frac{l_{\rm 3D}}{2} \\
\end{cases}
\end{equation}
where we assume there are $N$ elements on the half of the cylinder for a fair comparison with other geometries, however, the visible number is $N/2$ from Rx, i.e., on a quarter cylinder, due to the 3D structure. 
Besides, the area $S_3$ is calculated by 
\begin{equation}
S_3=\frac{\pi b l_{\rm 3D}}{2a}\sqrt{4a^2-l_{\rm 3D}^2}.
\end{equation}

We then employ the polar coordinates, denoting $x=a\cos\theta$ and $y=b\sin\theta$, and thus the area of $S_2$ can be calculated by
\begin{equation}
\begin{aligned}
S_2&=4\int_{\arccos\left(\frac{\pi l_{\rm 3D}}{4a}\right)}^{\frac{\pi}{2}} ab \sin\theta \cos\theta d\theta,\\
&=2ab\cos^2\left(\arccos\left(\frac{\pi l_{\rm 3D}}{4a}\right)\right),
\end{aligned}
\end{equation}
where substituting $a=\frac{l_{\rm 3D}}{2}$, it can be seen that $S_2=\frac{\pi^2 b l_{\rm 3D}}{4}=\pi a^{*}b$ and $S_3=0$, which proves the continuity. $S_2$ and $S_3$ are shown in Fig.~A1(b) with the explanation in Appendix B.
\begin{figure}[!t]
  \centering
   \subfigure[$N_{\rm eff}$ vs. $x_s$]{\includegraphics[width=1.7in]{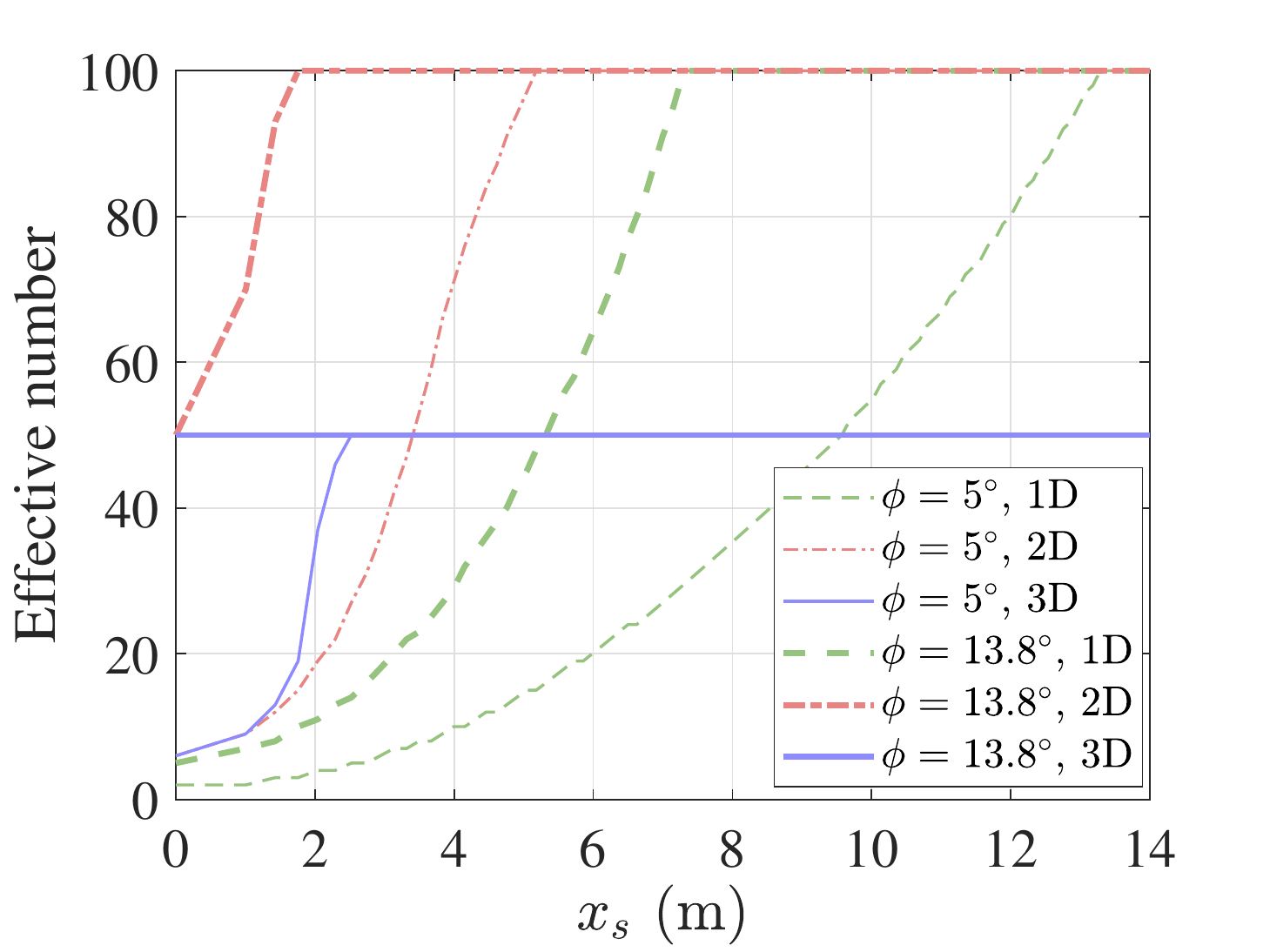}}
   \subfigure[$N_{\rm eff}$ vs. $\phi$]{\includegraphics[width=1.7in]{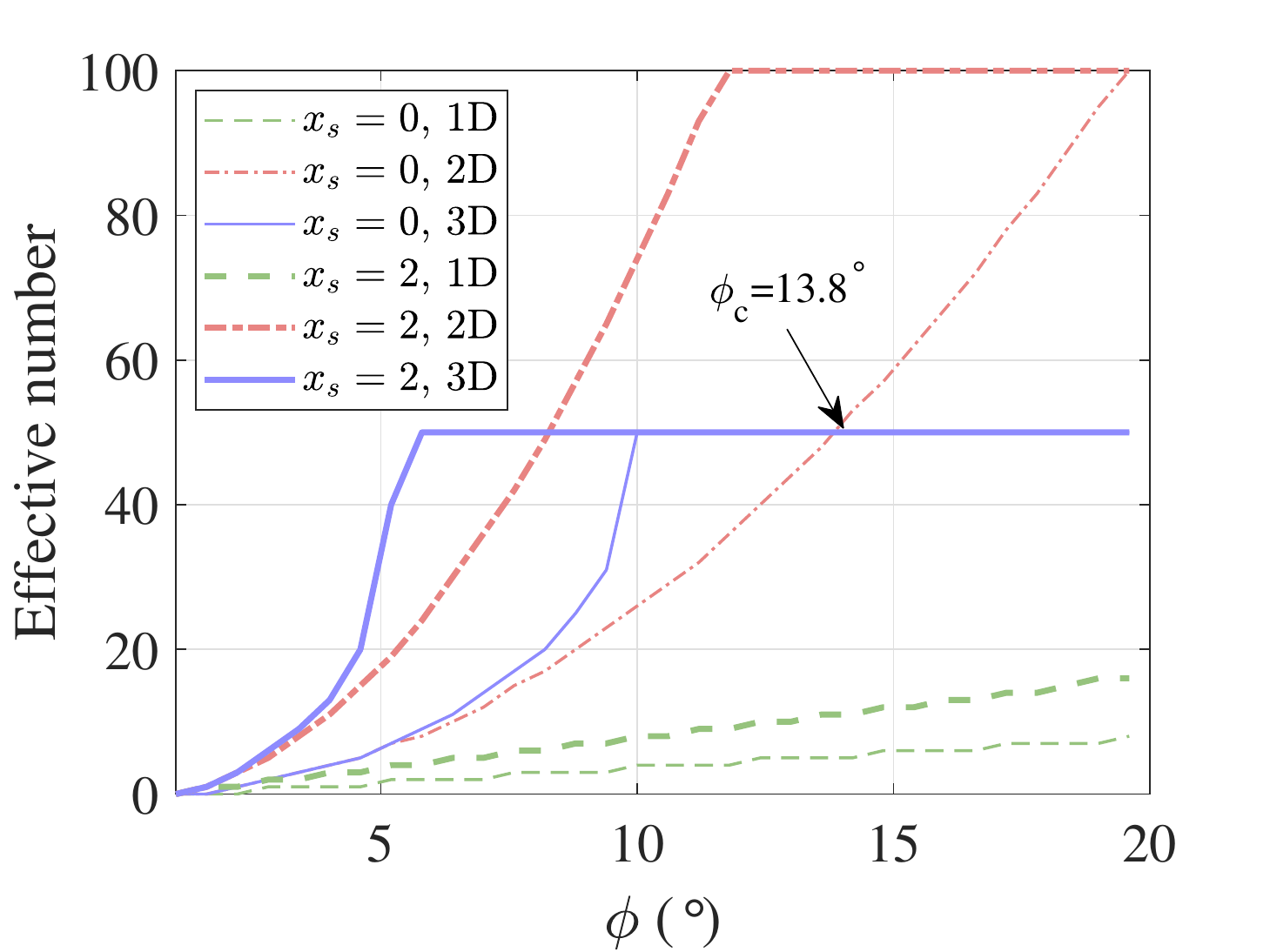}}
  \caption{Comparisons of effective numbers for different geometries where $y_s=2$~m, $h_t=h_s=3$~m: (a) $N_{\rm eff}$ versus $x_s$ ($x$-coordinate of RIS), (b) $N_{\rm eff}$ versus $\phi$, where we can find that for $x_s=0$ m, there is $\phi_c=13.8^\circ$ suggesting the critical beam-width, and for $\phi \ge \phi_c$, it exists $N_{\rm eff}^{\rm 2D}\ge N_{\rm eff}^{\rm 3D}$.}
  \label{eff_number}
 \end{figure}
\subsection{Fraunhofer Distance for Different RIS Geometries}
For a fair comparison, each RIS geometry is assumed to have the same number of elements $N$ distributed with the same spacing $d_s$ on the surface, giving the same element density on the surface for each geometry. 
The Fraunhofer distance used to determine near or far field for the 1D RIS is expressed as $d_{\rm 1D}=\frac{2D^2}{\lambda}=\frac{2(( N_{\rm eff}-1)d_s+ N_{\rm eff}d_x)^2}{\lambda}$, where $D$ is the largest dimension of the RIS \cite{tang21}.
We use $d_x = d_y=d_s=\frac{\lambda}{2}$, 
which yields $d_{\rm 1D}=\frac{\lambda}{2}(2N_{\rm eff}-1)^2$. 
The Fraunhofer distance for the 2D square RIS can be expressed $d_{\rm 2D}=\lambda(2\sqrt{N_{\rm eff}}-1)^2$ for fully illuminated condition where $D=\sqrt{2}l_{\rm 2D}$. For partial illumination, $d_{\rm 2D}=\frac{2(2a)^2}{\lambda}$ where $D=2a$.
Similarly, the Fraunhofer distance for the 3D RIS is then expressed as $d_{\rm 3D}=\lambda\left(\frac{\pi}{4}+\frac{1}{\pi}\right)N_{\rm eff}$ for fully illuminated condition where $D^2=(1+\frac{\pi^2}{4})l_{\rm 3D}^2$ and $d_{\rm 3D}=\frac{2(2a^*)^2}{\lambda}$ where $D=2a^*$ for partial illumination. For instance, in an indoor environment at 3.5~GHz with Tx-RIS center distance of $r_1$ (2-10~m), Fig.~\ref{far-field-distance} suggests that $d_{\rm 1D}$ increases faster than the others, because it is as function of $N_{\rm eff}^2$. Moreover, the 1D and 2D RISs experience more near fields than the 3D RIS since the more compact 3D structure leads to a smaller size of RIS. Moreover, when $N_{\rm eff}$ reaches the limit of each geometry, $d_G$ will become constant.

 \subsection{Tx-RIS and RIS-Rx Distances under Near Field}
For the near-field propagation, the assumption of the planar wave does not hold, which results in the distances dependent on the actual locations of the RIS elements. To determine $r_{1,n}$ and $r_{2,n}$ for $n=1...N_{\rm eff}$, the calculation is diverse for different RIS geometries. For the linear RIS, $r_{1,n}$ can be approximately calculated by the Law of cosines \cite{cuim22}, i.e., $r_{1,n}=\sqrt{r_1^2+(nd_s)^2+2r_1 nd_s\cos(\varphi_t)}$,
for $n=[-\frac{N_{\rm eff}}{2}, \frac{N_{\rm eff}}{2}]$ with $N_{\rm eff}$ numbers. The same method applies in $r_{2,n}$. We denote the coordinates of Tx, the center of RIS, and Rx as $(0,0,h_t)$, $(x_s,y_s,h_s)$, and $(x_r,y_r,h_r)$, respectively. 
For the 2D and 3D RISs, $r_{1,m,n}$ and $r_{2,m,n}$ can be written by $r_{1,m,n}=\sqrt{x_{m,n}^2+y_{m,n}^2+(z_{m,n}-h_t)^2}$ and $r_{2,m,n}=\sqrt{(x_{m,n}-x_r)^2+(y_{m,n}-y_r)^2+(z_{m,n}-h_r)^2}$, where it then yields the angular information, i.e., $\varphi_{t,m,n}=\arctan(x_{m,n}/y_s)$, $\theta_{t,m,n}=\arctan((z_{m,n}-h_t)/d_{1,m,n})$, $\varphi_{r,m,n}=\arctan((x_{m,n}-x_r)/(y_{m,n}-y_r))$, and $\theta_{r,m,n}=\arctan((z_{m,n}-h_r)/d_{2,m,n})$. We consider the following conditions to express the coordinates of RIS elements:
\subsubsection{Fully illuminated} The locations of elements in the 2D RIS can be expressed as ($x_{m,n}$, $y_{m,n}$, $z_{m,n})=(x_s+nd_s$, $y_s$, $h_s+md_s$) for $m,n=[-\frac{\sqrt{N}}{2}, \frac{\sqrt{N}}{2}]$ with $\sqrt{N}$ number. For the 3D RIS, it exists ($x_{p,q}$, $y_{p,q}$, $z_{p,q})=(x_s+\frac{pd_s^{'}}{\cos(pd_s/l_{\rm3D})}$, $y_s+\frac{pd_s^{'}}{\cos(pd_s/l_{\rm3D})}$, $h_s+ q d_s$) where $q=[-\frac{l_{\rm3D}}{2(d_x+d_s)},\frac{l_{\rm3D}}{2(d_x+d_s)}]$ with $\frac{l_{\rm3D}}{d_x+d_s}$ numbers, and $p=[-\frac{N(d_x+d_s)}{4l_{\rm3D}},\frac{N(d_x+d_s)}{4l_{\rm3D}}]$ with $\frac{N(d_x+d_s)}{2l_{\rm3D}}$ numbers, and $d_s^{'}=\sqrt{\frac{l_{\rm3D}^2}{2}-\frac{l_{\rm3D}^2}{2}\cos(2d_s/l_{\rm3D})}$ is the chord length of the arc length $d_s$, based on the Law of cosines. 
\subsubsection{Partially illuminated} The number of element in the major and minor axes of an ellipse can be expressed as $N_a=a/d_s$ and $N_b=b/d_s$. For the total illuminated number $N_{\rm eff}$, the indexes $(m,n)$ and $(p,q)$ of elements in each column and row can be determined. Then, the coordinates ($x_{m,n}$, $y_{m,n}$, $z_{m,n}$) and ($x_{p,q}$, $y_{p,q}$, $z_{p,q}$) will be selected from the coordinates obtained in the fully illuminated condition.
\begin{figure}[!t]
  \centering
 \includegraphics[width=2.8in]{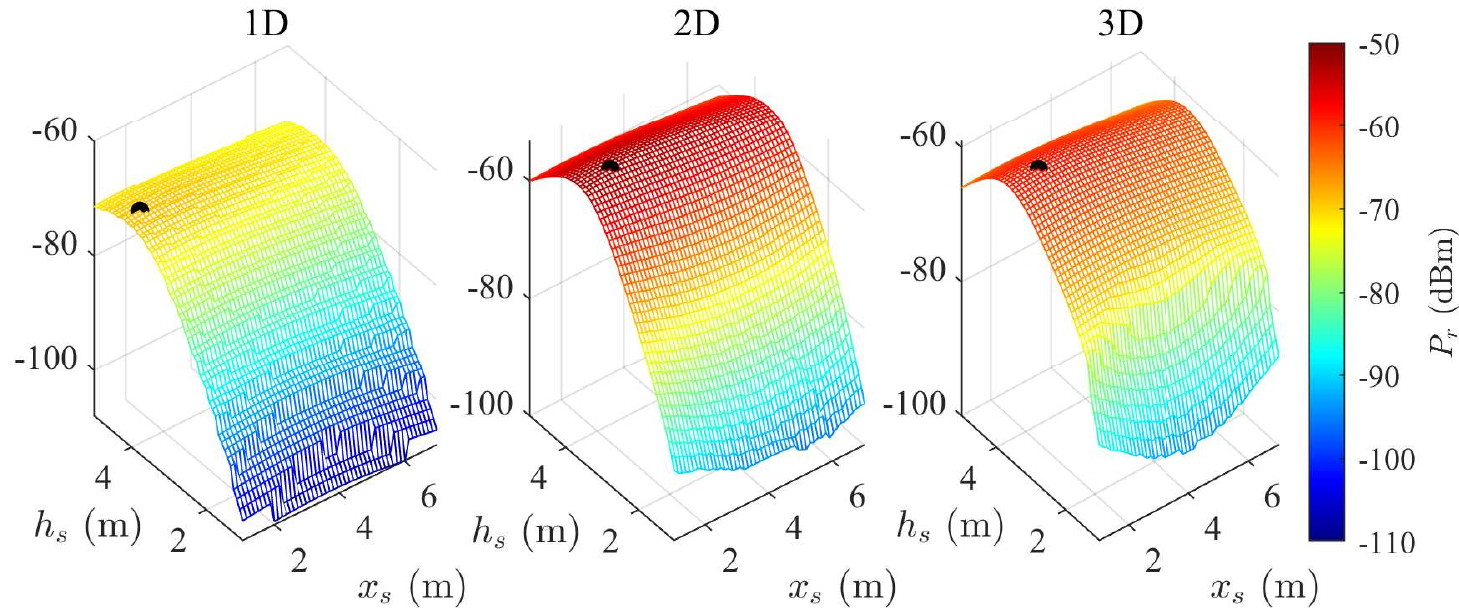}
\caption{Received power $P_r$ versus $x_s$ and $h_s$ for different geometries where $P_t=0$~dBm, $\phi=5^\circ$, $(x_r,y_r,h_r)=(5,0,1.5)$~m, $h_t=$3~m, and $y_s=$2~m.}
 \label{rp_ris}
 \end{figure} 
\begin{figure}[!t]
  \centering
   \subfigure[]{\includegraphics[width=1.7in]{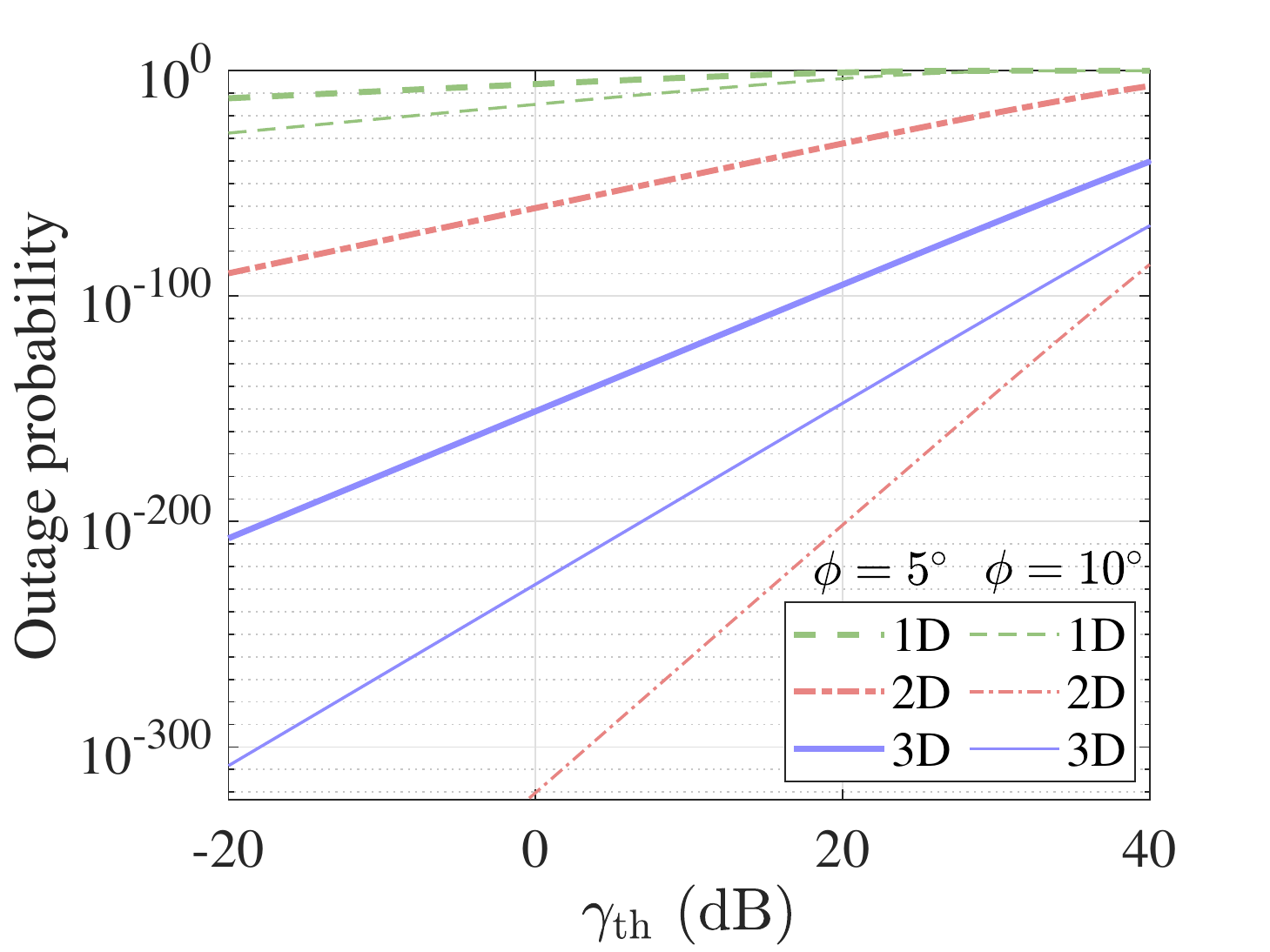}}
    \subfigure[]{\includegraphics[width=1.7in]{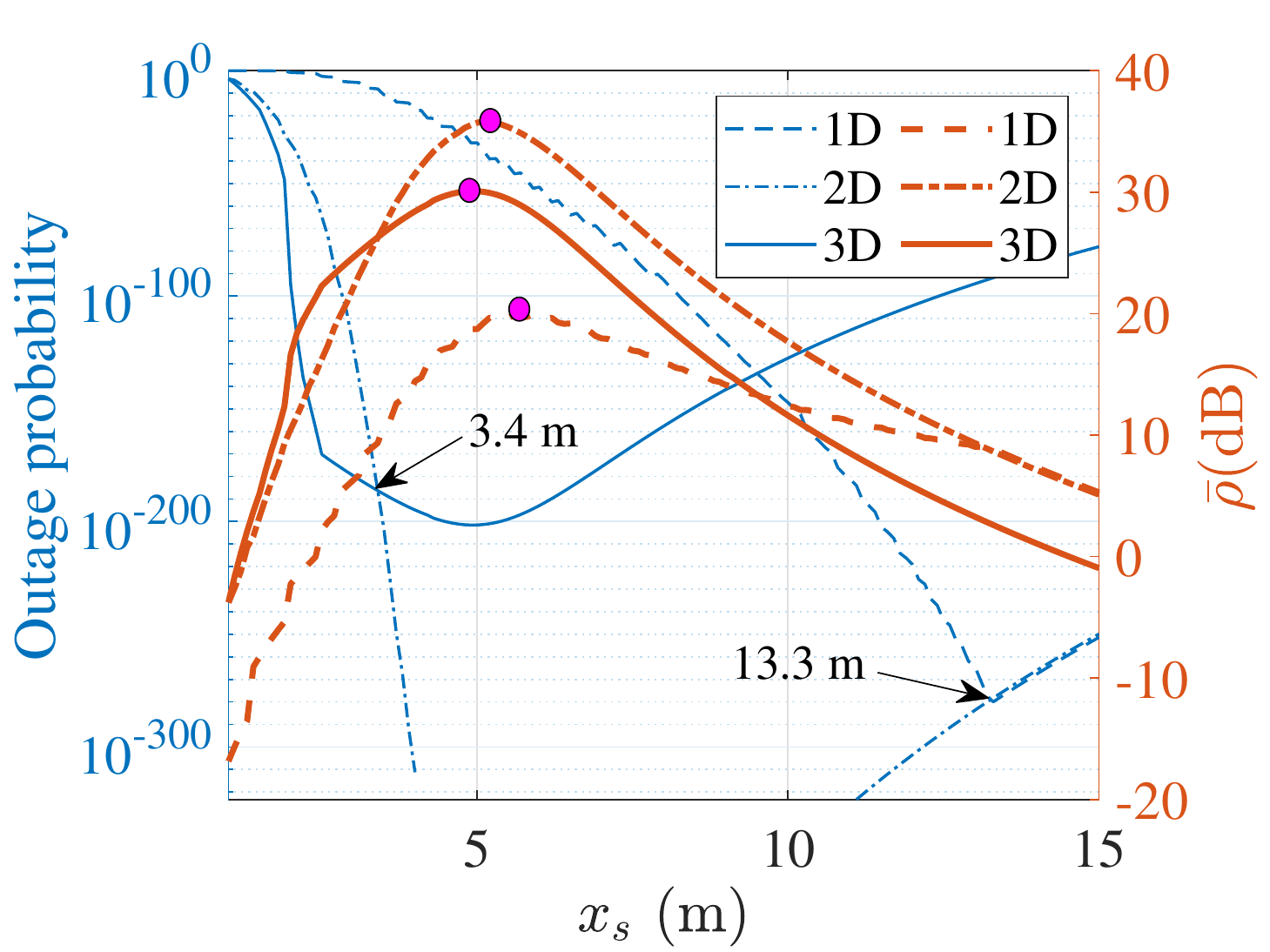}}
  \caption{Outage probability and SNR: (a) $P_{\rm out}$ vs. $\gamma_{\rm th}$ for $\phi=5^\circ$/$10^\circ$, and (b) $P_{\rm out}$ and $\bar{\rho}$ vs. $x_s$ for $(x_r,y_r,h_r)=(5,0,1.5)$~m, and $\gamma_{\rm th}=20$~dB. }
  \label{op_snr}
 \end{figure} 
\subsection{Outage Probability}
The outage probability is a popular metric for evaluating the performance, which is defined as the probability that the SNR is smaller than a specific threshold $\rho_{\rm th}$, expressed by
\begin{equation}
P_{\rm out}(\rho_{\rm th})=\mathbf{Prob}(\rho_{\max} \le \rho_{\rm th})=F_{\rho_{\max}}(\rho_{\rm th}),
\end{equation}
where $F_{\rho_{\max}}(\cdot)$ denotes cumulative distribution function (CDF) of $\rho_{\max}$. 
With the moment-matching method \cite{cui21}, it can be obtained as $F_{\rho_{\max}}(\rho_{\rm th})=\frac{\Gamma\left(kN_{\rm eff},\delta\sqrt{\frac{\rho_{\rm th}}{\bar{\rho}}}\right)}{\Gamma(kN_{\rm eff})}$,
where $\Gamma(\cdot,\cdot)$ is the lower incomplete Gamma function, $\Gamma(\cdot)$ is the Gamma function, and $k=\frac{\pi^2}{16-\pi^2}$, $\delta=\frac{2\pi}{(16-\pi^2)\sigma}$ with $\sigma=\sigma_1\sigma_2$.

Denote $\eta=\alpha\beta$, for an arbitrary number of $N_{\rm eff}$, $\eta$ can be approximated as Gamma distribution, there is $\eta\sim\mathcal{G}(k,\delta)$. According to the addictive characteristic of Gamma distribution, we obtain that $A \sim \mathcal{G}(kN_{\rm eff},\delta)$. Correspondingly, we can obtain that $A^2$ follows the generalized Gamma distribution, whose CDF can be expressed as 
\begin{equation}
F_{A^2}(x)=\frac{\Gamma\left(kN_{\rm eff},\frac{\sqrt{x}}{\delta}\right)}{\Gamma(kN_{\rm eff})},
\end{equation}
where the outage probability can be obtained by substituting $x=\frac{\rho_{\rm th}}{\bar{\rho}}$. Detailed derivations can refer to our prior work \cite{cui21}.  
 

\section{Numerical Results}
In this section, we will discuss numerical results. The following parameters are used: $N=100$, $P_t=0$~dBm, $P_n=-100$~dBm, $f=3.5$~GHz, $(x_t,y_t,h_t)=(0,0,3)$~m, $(x_r,y_r,h_r)=(5,0,1.5)$~m, $y_s=2$~m, $\sigma=1$, and $\phi=5/10^\circ$. Some parameters may change for different analytical purposes. 
 \subsection{Effective Number and Received Power}
We first illustrate the relationship between $N_{\rm eff}$ and $x_s$ with fixed HPBW $\phi$, as well as $N_{\rm eff}$ and $\phi$ with fixed $x_s$ in Fig.~\ref{eff_number}. The slope of the results demonstrates that $N_{\rm eff}$ increases faster for compact geometries, hence faster for 
3D, 2D, and 1D RIS, respectively
. Moreover, for a short distance and narrow HPBW, resulting in the smallest illumination area, the 3D RIS has the highest $N_{\rm eff}$. 
However, 
by increasing $x_s$ and $\phi$, the 2D and even 1D RIS become more effective. 
This is mainly caused by the fact that the circular geometry directs some RIS elements away from the Tx and Rx.  
When increasing $x_s$ and $\phi$, all visible elements will contribute for all geometries, resulting in maximal performance for both the 1D and 2D RIS given by $N_{\rm eff}=N$. The 3D RIS suffers from reduced visibility, and in our placement assumptions only half of the elements contribute and $N_{\rm eff}=N/2$. 
Finally, the critical HPBW $\phi_c$ in Fig.~\ref{eff_number}(b) verifies the correctness of our theoretical derivation. 

To analyze the received power using different RISs, we change both horizontal and vertical locations of the RIS. As shown in Fig.~\ref{rp_ris}, it is found that the 2D RIS allows a higher received power than the other RISs. Moreover, the optimal RIS location ($x_s^{*}$, $h_s^{*}$) for the 1D RIS is (1.49, 4.18)~m, while the 2D and 3D RISs have the same optimal location, i.e., (1.98, 3.61)~m, while the maximum power is -69~dBm, -53~dBm, and -59~dBm for the 1D, 2D, and 3D RIS, respectively. Overall, the received power for fixed $x_s$ increases for $h_s \le h_s^{*}$ and then decreases for a higher $h_s$, due to the heights impacting $r_1$, $r_2$ and $P_r \sim \frac{1}{r_1^2r_2^2}$. While for fixed $h_s$, the received power presents different trends along $x_s$, since the received power is determined by the joint effect of $N_{\rm eff}$, $r_1$, $r_2$, and $\varphi_t,\varphi_r$ based on Eq.~(\ref{pr_near}). Nonetheless, the optimal RIS location for given Tx and Rx locations, and HPBW, can be obtained.



\subsection{SNR and Outage Probability}

We then indicate the outage probability and SNR results considering the fixed and horizontally moving location of RIS and combining the effect of path loss and small scale fading. Fig.~\ref{op_snr}(a) shows the outage probability for $\phi=5/10^\circ$ and the fixed RIS with $(x_s,y_s,h_s)=(2,2,3)$~m. 
The results show the 3D and 2D RISs are superior in performance for the small and large $\phi$, respectively. 
For the moving RIS, we show $P_{\rm out}$ and mean SNR $\bar{\rho}$ simultaneously in Fig.~\ref{op_snr}(b). It is found that the maximum average SNR is obtained at 
$x_s^{\rm opt}$ = 4.9, 5.1, and 5.6~m with SNR values of 29.9, 35.8, and 20.2~dB for 3D, 2D, and 1D RIS, respectively, where geometric compactness leads to the difference of $x_s^{\rm opt}$ and the 2D RIS can achieve the highest average SNR. 
Besides, the outage probabilities of 2D and 3D RISs are inversely proportional to the SNR, while $P_{\rm out}$ decreases with a declined SNR for the 1D RIS in $x_s\le 13.3$~m, which is because $P_{\rm out}$ is jointly determined by $N_{\rm eff}$ and $\bar{\rho}$, while $N_{\rm eff}^{\rm 1D}$ is increasing in this range. When it reaches the limit number, $P_{\rm out}$ increases with a decreasing $\bar{\rho}$. Both $P_{\rm out}$ and $\bar{\rho}$ results indicate the suitable deploying range for 3D and 2D is separated at $x_s=3.4$~m. Besides, the same performance can be obtained for the 1D and 2D RISs when $x_s\ge13.3$~m.




\section{Conclusions}
In this paper, we presented a general analytical framework for the impact of geometric structures of RIS on communication performances, under the high-directional transmitting beam. To obtain the illuminated area, effective number, near-field distances, and outage probability, we employed geometry-based and moment-matching-based analytical methods. In general, analytical results show that the 3D and 2D RISs can be respectively used for the shorter/longer distance and narrower/wider beam-width, e.g., the horizontal distance $x_s=3.4$~m and the HPBW $\phi=13.8^\circ$ considered as cut-off points for an indoor environment. 
Besides, for reducing near-field propagation effects, the 3D RIS is a better option to deploy than the others, thanks to its higher structural compactness. 


\appendix
\renewcommand\thefigure{\Alph{section}\arabic{figure}}    \setcounter{figure}{0}   
\begin{figure}[!t]
  \centering
   \subfigure[]{\includegraphics[width=1.3in]{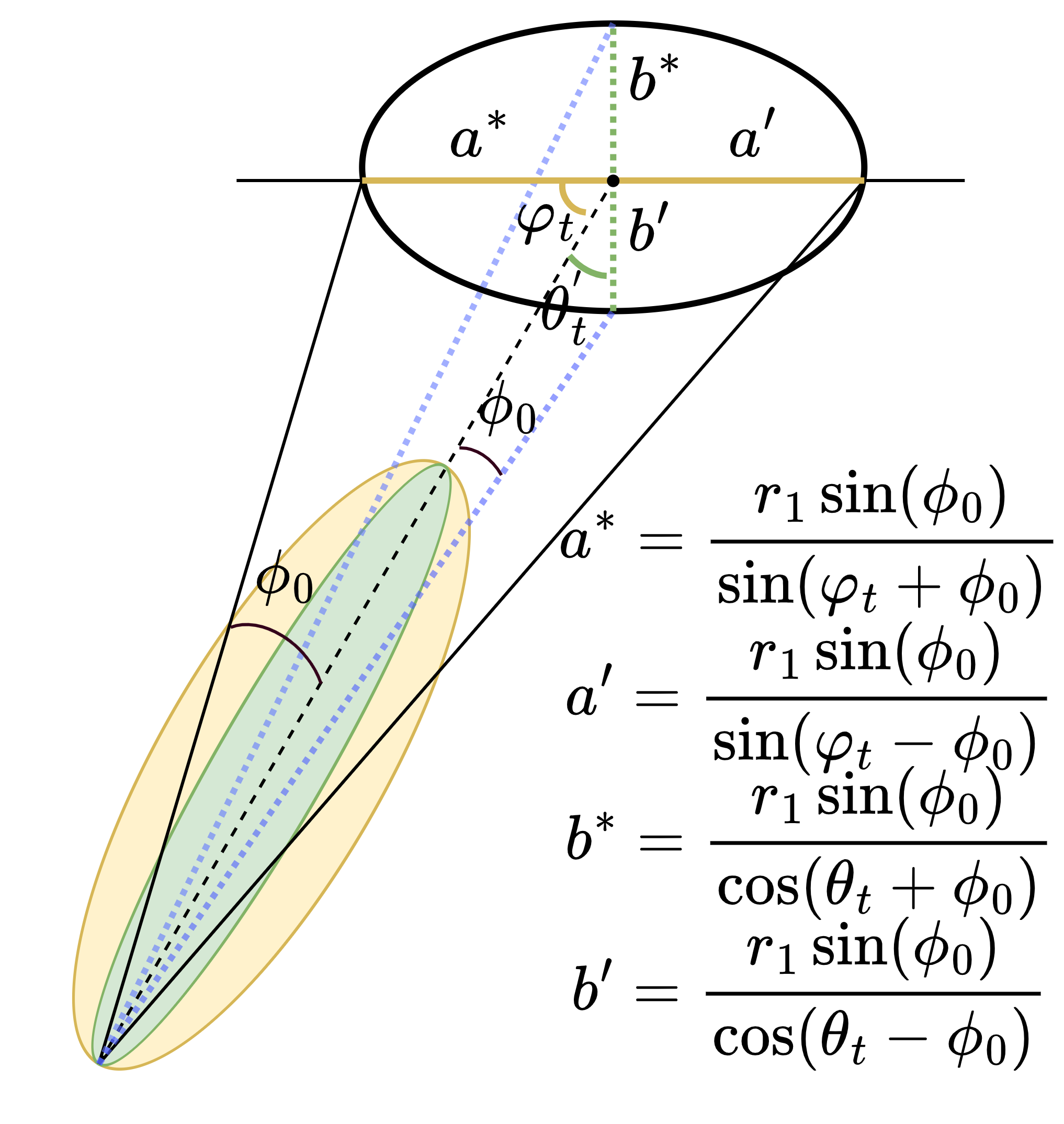}}
    \subfigure[]{\includegraphics[width=1.9in]{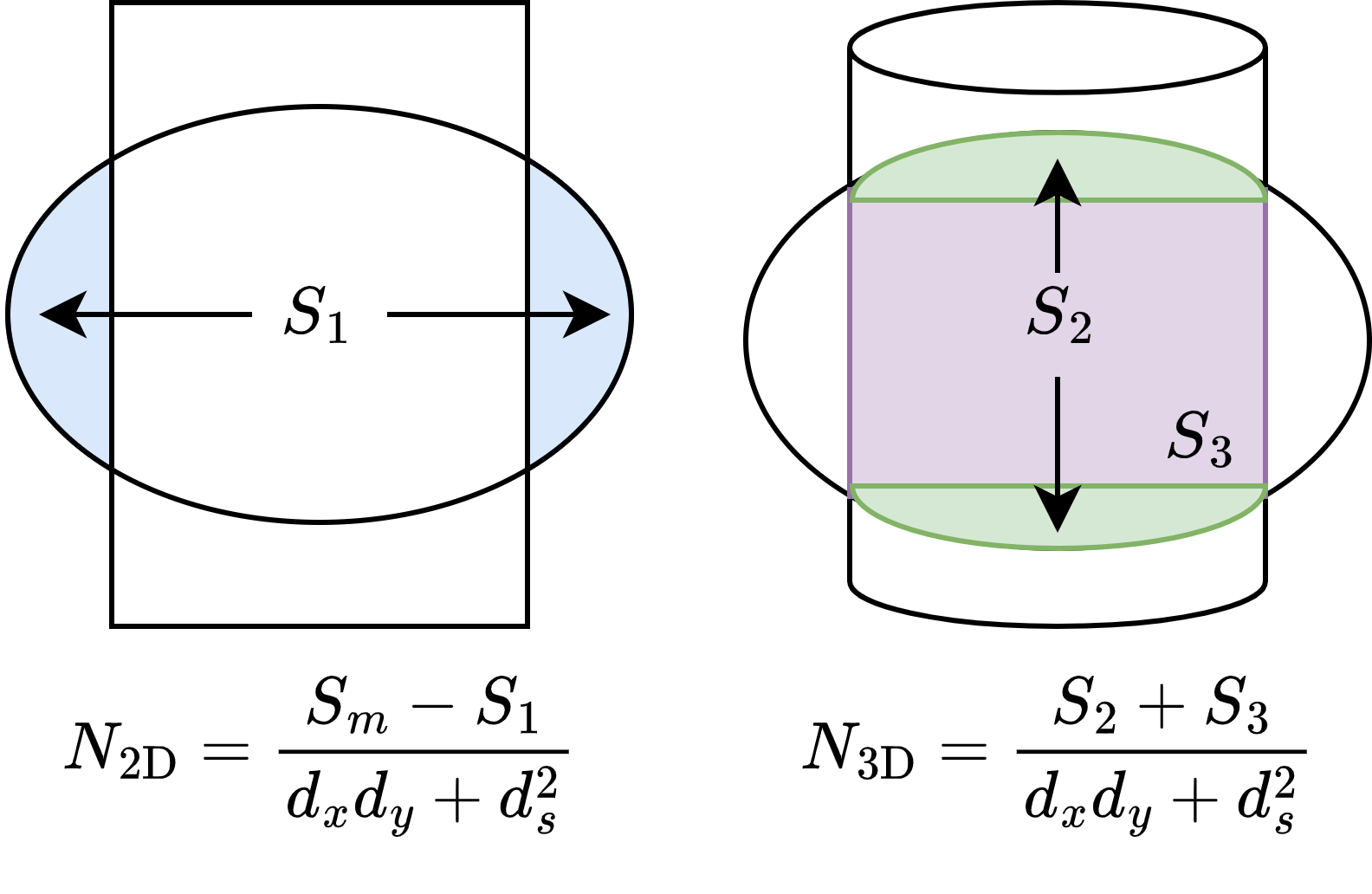}}
  \caption{Geometric calculation: (a) Calculation of the major and minor axes for the illuminated area, (b) Area Calculation for $a > \frac{l_{\rm 2D,3D}}{2},b < \frac{l_{\rm 2D,3D}}{2}$.}
  \label{fig3_calculation}
 \end{figure}
\subsection{Graphical Interpretation of Illuminated Area Calculation}
Consider the 3D deployment of RIS, the beam's projection on the RIS is influenced by two departure angles from the Tx, i.e., AoD ($\varphi_t$)  and EoD ($\theta_t$). The area is structured by four lengths, i.e., $a^{'}$, $a^{*}$, $b^{'}$, and $b^{*}$. Based on the Law of Sines, for instance, $\frac{a^*}{\sin(\phi_0)}=\frac{r_1}{\sin(\varphi_t+\phi_0)}$, these values are given in Fig.~A1(a). Note that $\theta_t^{'}=\frac{\pi}{2}-\theta_t$ in the figure. However, the shape of illuminated area is irregular due to the 3D departure. Specifically, there are $a^{'}\neq a^{*}$ and $b^{'}\neq b^{*}$ for most cases. However, when the Tx and RIS are at the same height, the actual shape is an asymmetric oval ($a^{'}\neq a^{*}$ and $b^{'}=b^{*}$). Herein, we approximately employ an ellipse to express the illuminated area, with the average major axis $a=\frac{1}{2}(a^{'}+a^{*})$ and the average minor axis $b=\frac{1}{2}(b^{'}+b^{*})$ by assuming the larger part of the shape filling up the smaller part for average.
\subsection{Graphical Interpretation of Effective Number Calculation}
For the 2D RIS, as shown in Fig.~A1(b), we show the most challenging condition, i.e., $a > \frac{l_{\rm 2D}}{2},b < \frac{l_{\rm 2D}}{2}$, where $S_1$ can be obtained by the integral of the elliptic Equation, and then the illuminated area can be expressed as $S_{\rm m}-S_1$ where $S_1$ is given in Eq.~(7). It is also a challenging condition for the 3D RIS. However, because of the structure of the cylinder, we separate the illuminated area into $S_2+S_3$, which can be obtained by the integral operation using the polar coordinate, where $S_2$ and $S_3$ are given in Eq.~(10) and Eq.~(11), respectively.
\ifCLASSOPTIONcaptionsoff
  \newpage
\fi


\begin{thebibliography}{99}


\bibitem{Strinati21}
E. C. Strinati \emph{et al.}, ``Reconfigurable, intelligent, and sustainable wireless environments for 6G smart connectivity," \emph{IEEE Commun. Mag.}, vol. 59, no. 10, pp. 99-105, Oct. 2021.


\bibitem{cui21} 
Z. Cui, K. Guan, J. Zhang and Z. Zhong, ``SNR coverage probability analysis of RIS-aided communication systems,'' \emph{IEEE Trans. Veh. Technol.}, vol. 70, no. 4, pp. 3914-3919, Apr. 2021.

\bibitem{salhab21}
A. M. Salhab and M. H. Samuh, ``Accurate performance analysis of reconfigurable intelligent surfaces over Rician fading channels," \emph{IEEE Wireless Commun. Lett.}, vol. 10, no. 5, pp. 1051-1055, May 2021.

\bibitem{hhu21}
H. Ibrahim, H. Tabassum and U. T. Nguyen, ``Exact coverage analysis of intelligent reflecting surfaces with Nakagami-\emph{M} channels," \emph{IEEE Trans. Veh. Technol.}, vol. 70, no. 1, pp. 1072-1076, Jan. 2021.

\bibitem{dai20} 
L. Dai \emph{et al.}, ``Reconfigurable intelligent surface-based wireless communications: Antenna design, prototyping, and experimental results," \emph{IEEE Access}, vol. 8, pp. 45913-45923, Mar. 2020.

\bibitem{basar21}
E. Basar, I. Yildirim and F. Kilinc, ``Indoor and outdoor physical channel modeling and efficient positioning for reconfigurable intelligent surfaces in mmWave bands," \emph{IEEE Trans. Commun.}, vol. 69, no. 12, pp. 8600-8611, Dec. 2021.

\bibitem{JQP22}
Z. Ji, Z. Qin and C. G. Parini, ``Reconfigurable intelligent surface aided cellular networks with device-to-device users," \emph{IEEE Trans. Commun.}, vol. 70, no. 3, pp. 1808-1819, Mar. 2022.


\bibitem{ammc22} 
A. Al-Hilo, M. Samir, M. Elhattab, C. Assi and S. Sharafeddine, ``Reconfigurable intelligent surface enabled vehicular communication: Joint user scheduling and passive beamforming," \emph{IEEE Trans. Veh. Technol.}, vol. 71, no. 3, pp. 2333-2345, Mar. 2022.




\bibitem{huang19} 
C. Huang, A. Zappone, G. C. Alexandropoulos, M. Debbah and C. Yuen, ``Reconfigurable intelligent surfaces for energy efficiency in wireless communication," \emph{IEEE Wireless Commun.}, vol. 18, no. 8, pp. 4157-4170, Aug. 2019.


\bibitem{basar19} 
E. Basar, M. Di Renzo, J. De Rosny, M. Debbah, M. -S. Alouini and R. Zhang, ``Wireless communications through reconfigurable intelligent surfaces," \emph{IEEE Access}, vol. 7, pp. 116753-116773, Aug. 2019.



\bibitem{Ntontin2021}
K. Ntontin \emph{et al.}, ``Optimal reconfigurable intelligent surface placement in millimeter-Wave communications,'' in \emph{Proc. EuCAP}, Virtual, Mar. 2021, pp. 1-5.

\bibitem{Ntontin2022}
K. Ntontin, A. -A. A. Boulogeorgos, D. G. Selimis, F. I. Lazarakis, A. Alexiou and S. Chatzinotas, ``Reconfigurable intelligent surface optimal placement in millimeter-wave networks,'' \emph{IEEE Open J. Commun. Soc.}, vol. 2, pp. 704-718, Mar. 2021.

\bibitem{GSA2022}
G. Stratidakis, S. Droulias and A. Alexiou, ``Understanding the RIS efficiency: From partial to full illumination,'' \emph{Proc. IEEE 23rd SPAWC}, Oulu, Finland, Jul. 2022, pp. 1-5.

\bibitem{yzh17}
X. Yu, J. Zhang, M. Haenggi and K. B. Letaief, ``Coverage analysis for millimeter wave networks: The impact of directional antenna arrays,'' \emph{IEEE J. Sel. Areas Commun.}, vol. 35, no. 7, pp. 1498-1512, Jul. 2017.

\bibitem{tang21}
W. Tang \emph{et al.}, ``Wireless communications with reconfigurable intelligent surface: Path loss modeling and experimental measurement," \emph{IEEE Trans. Wireless Commun.}, vol. 20, no. 1, pp. 421-439, Jan. 2021.

\bibitem{cuim22}
M. Cui and L. Dai, ``Channel estimation for extremely large-scale MIMO: Far-field or near-field?," \emph{IEEE Trans. Commun.}, vol. 70, no. 4, pp. 2663-2677, Apr. 2022.



\end{thebibliography}
\end{document}